\documentclass[final]{aipproc}
\layoutstyle{6x9}

\newcommand{\unit}[1]{\ifmmode {\rm\ #1} \else {$\rm #1$} \fi}

\newcommand{\SPEAR}{{\it SPEAR}}
\newcommand{\OIIIb}{{\rm O{\sc iii}]}}
\newcommand{\OIV}{{\rm [O{\sc iv}]}}
\newcommand{\OVI}{{\rm O{\sc vi}}}
\newcommand{\HeII}{{\rm He{\sc ii}}}
\newcommand{\CIII}{{\rm C{\sc iii}}}
\newcommand{\CIV}{{\rm C{\sc iv}}}
\newcommand{\SiIV}{{\rm Si{\sc iv}}}

\newcommand{\solarmassesperyr}{\unit{M_\odot yr^{-1}}}

\newcommand{\angstrom}{\AA}
\newcommand{\arcmin}{\unit{^\prime}}
\newcommand{\pressure}{\unit{cm^{-3} K}}
\newcommand{\lu}{\unit{photon\ s^{-1} cm^{-2} sr^{-1}}}
\newcommand{\doublet}{\unit{\lambda\lambda}}
\newcommand{\singlet}{\unit{\lambda}}

\begin{document}

\title{Implications of the \SPEAR\ FUV Maps on Our Understanding of the ISM}


\author{Eric J Korpela}
{address={Space Sciences Laboratory, University of California, Berkeley, CA 94720 USA}}
\author{Martin Sirk}
{address={Space Sciences Laboratory, University of California, Berkeley, CA 94720 USA}}
\author{Jerry Edelstein}
{address={Space Sciences Laboratory, University of California, Berkeley, CA 94720 USA}}
\author{Kwangil Seon}
{address={Korea Astronomy and Space Science Institute, 305-348, Daejeon, Korea}}
\author{Kyoung-Wook Min}
{address={Korea Advanced Institute of Science and Technology, 305-701, Daejeon, Korea}}
\author{Wonyong Han}
{address={Korea Astronomy and Space Science Institute, 305-348, Daejeon, Korea}}

\begin{abstract}
The distribution of a low-density transition temperature ($10^{4.5}-10^{5.5}$ K)
gas in the interstellar medium conveys the character and evolution of diffuse
matter in the Galaxy. This difficult to observe component of the ISM emits
mainly in the far-ultraviolet (FUV) (912-1800 \angstrom) band. 
We describe spectral maps
of FUV emission lines from the highly ionized species \CIV\ and \OVI\ likely to
be the dominant cooling mechanisms of transition temperature gas in the ISM. The
maps were obtained using an orbital spectrometer, \SPEAR, that was
launched in 2003 and has observed the FUV sky with a spectral resolution of
$\sim 550$ and an angular resolution of 10\arcmin.

We compare distribution of flux in these maps with three basic models of the
distribution of transition temperature gas.  We find that the median
distribution of \CIV\ and \OVI\ emission is consistent with the spatial
distribution and line ratios expected from a McKee-Ostriker (MO) type model of
evaporative interfaces. However,
the intensities are a factor of three higher than would be expected at the MO
preferred parameters.  Some high intensity regions are clearly associated with
supernova remnants and superbubble structures.  Others may indicate regions 
where gas is cooling through the transition temperature.
\end{abstract}

\maketitle

\section{Introduction}

The far-UV (900-1800 \angstrom) spectrum of diffuse interstellar emission
contains astrophysically important cooling lines, among them the \OVI\ doublet
($\doublet 1032,1038$) represents the dominant radiative cooling
mechanism for gas with temperatures between 10$^{5.3}$ and 10$^{5.7}$
K. The \CIV\ line ($\doublet 1548,1551$) is an important radiative
cooling mechanism for 
gas with temperatures between 10$^{4.9}$ and 10$^{5.3}$ K \citep{chianti}.
Because of the high cooling rates due to these and other lines, gas
in this temperature range cools rapidly to lower temperatures, and
therefore we refer to gas in this temperature range as ``transition
temperature gas.''  Due to the rapid cooling, in order to be observed
this gas must be replenished, either from a source of higher temperature
gas cooling through this temperature range, shock heating of cooler gas,
conductive heating in a boundary between hot and cold gas, or turbulent
mixing of hot and cold gas \citep{mckost77, spitzer90, slavin93}.

The Spectroscopy of Plasma Evolution from Astrophysical Radiation 
(\SPEAR) instrument, 
designed for observing emission lines from the diffuse ISM,
was launched in late 2003.
\SPEAR\, is a dual-channel FUV imaging spectrograph
(short-$\lambda$ (S) channel 900 - 1150 \AA, 
long-$\lambda$ (L) channel 1350 - 1750 \AA)
with 
$\lambda/\Delta\lambda\sim$550,
with a large field of view (FOV)
(S: 4.0$^{\circ} \times$ 4.6', 
L: 7.5$^{\circ} \times$ 4.3')
imaged at 10$\arcmin$ resolution.
See \cite{edelstein06a} and \cite{edelstein06b} for an overview of the
instrument and mission, and \cite{korpela06} for a discussion of data analysis.
\SPEAR\ sky-survey observations consisted of sweeps at constant ecliptic
longitude from the
north ecliptic pole to the south ecliptic pole through the 
anti-solar point.  Throughout the course of the mission about 70\% of the
sky was observed.

\section{Basic properties of FUV radiative cooling}

When modeling FUV emission from transition temperature gas it is vital to
consider non-equilibrium effects.  The cooling due to FUV line emission is
efficient enough that for optically thin gas the radiative cooling timescale 
is always short compared to the recombination timescales.  This is true
regardless of the pressure or density of the gas in question.  Fortunately there
are several rules of thumb that can be used to explain the behavior of line
emission as gas is heated or cools.  These derive from the fact that ionization
always lags the temperature.

When gas is heated the ionization state mimics that of a lower temperature gas,
therefore the FUV lines are generated in gas that is hotter than would be the
case for collisional ionization equilibrium (CIE) emission.  Heated gas typically 
expands, and therefore the
densities at which line emission occurs is lower than the CIE value,
which reduces the line intensity.  Since the intensity is proportional to $n^2
\propto \frac{P^2}{T^2}$, emission from low stage ions is typically reduced less 
than that of high stage ions.  In this case we expect both \OVI\ and \CIV\ 
intensities to be reduced and that the ratio of \OVI\ intensity to \CIV\ intensity 
will also be reduced.

When gas cools through the transition temperature the opposite is true.  The
emission occurs at lower temperatures and higher densities than is true in CIE.
This tends to increase intensities over the CIE values and boost high stage ion
emission relative to low stage ions.  Therefore, in this case we would expect
both \OVI\ and \CIV\ intensity to increase and for the ratio of \OVI\ intensity to
\CIV\ intensity to increase.

The exact details of how these intensities and ratios change is determined by
physical factors, whether the heating is constant or impulsive, whether the
pressure is being maintained externally and whether the local pressure can 
change rapidly enough to match the temperature changes.  But in general these
rules of thumb appear to hold. 

\section{Basic Models of FUV emission}

We consider three simple models of the state of hot and intermediate temperature
gas in our Galaxy: 1) A McKee-Ostriker type model of neutral clouds evaporating
in a pervasive hot ($>10^6$ K) medium \citep{mckost77}. 2) A Galactic fountain in
which hot ($>10^6$ K) gas rises buoyantly above the Galactic plane where it
cools and condenses into clouds which fall back into the Galactic plane
\citep{spitzer90}. 3) Isolated supernova remnants distributed in an exponential
disk evolving in a non-clumpy medium \citep{slavin93}.  In each case a
non-equilibrium plasma emission model was utilized to calculate the expected
emission.  More details of these models can be found in \cite{korpela97}.

\begin{figure}[tbp]
\includegraphics[width=6in]{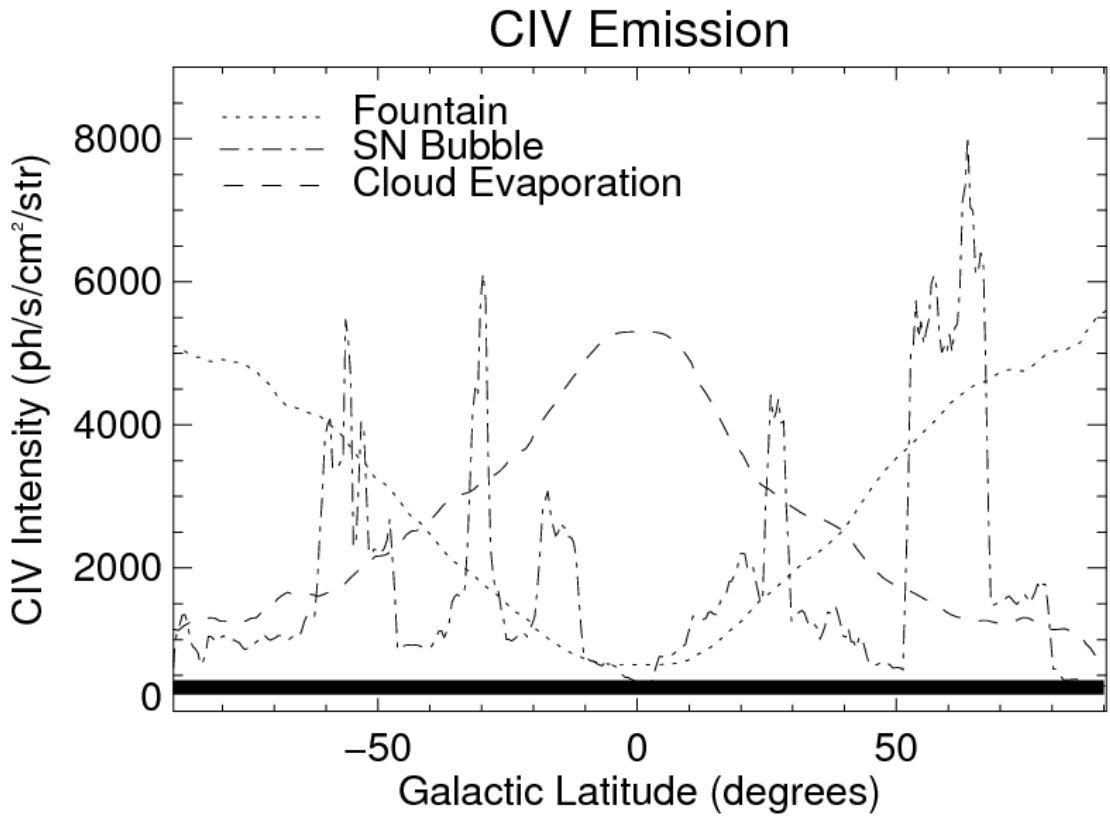}
\caption{The distribution of average \CIV\ intensity vs galactic latitude for
the three models discussed in the text.\label{modsvsb}}
\end{figure}

\noindent
{\bf McKee-Ostriker Model:}  The predominant emission in this model is
located in conductive cloud boundaries where cool gas is being heated by contact
with the hot medium.  Because of this the emitting regions are well
correlated with the neutral hydrogen clouds they surround. Therefore the
emission is brighter in the Galactic plane.  Dust opacity limits the maximum 
intensity seen and sets the ratio between plane and pole intensities.  Because
the opacity is higher at \OVI, \OVI\ shows a smaller plane to pole ratio.  For the
``preferred'' MO pressure of 2500 \pressure, the Galactic plane intensity is
5300 \lu (hereafter LU) in \CIV\ and 1600 LU in \OVI.  The distribution of
emission vs Galactic latitude in shown in Figure~\ref{modsvsb} as the dashed
line.
\\\ \\
\noindent
{\bf Galactic Fountain Model:}  In this model the hot gas is presumed to rise to
many times the scale height of the neutral Galactic gas where it cools
radiatively and falls back into the galactic plane.  Because of
instabilities the cooling gas is expected to form clumps of higher density where
the emission will be more intense.
The intensity of the emission depends upon the rate at which gas cools through
the transition temperature (which is the same as the total mass flow through the
fountain), and the ratio of \OVI\ to \CIV\ emission depends
upon the initial temperature of the hot gas and the pressure evolution of the gas 
as it cools.  We have ignored the possibility of turbulent mixing of hot and
cold gas in this model. For initial temperatures above 10$^6$ K, \OVI\
is typically expected to be a factor of 2 to 10 brighter than \CIV.  The model
shown is a high scale height model, scaled to match Galactic pole \CIV\ emission 
of 5000 LU, which corresponds to a mass flow rate of between 6 and 22
\solarmassesperyr. This is intended to be illustrative of the distribution on
the sky, as \CIV\ and \OVI\ measurements put upperlimits to the mass flow in a
hot galactic fountain below this level. 
\\\ \\
\noindent
{\bf Isolated SNR}  In this model FUV emission is concentrated into limb
brightened SNR.  The intrinsic intensity and line ratios of any remnant is
dependent upon the age of the remnant and the pressure and density of the medium
into which it is expanding.  The Galactic plane appears dark in this model
because the absorbing dust is presumed to be uniform outside of the remnants.
The average intensity at a given $b$ is likely to
be dominated by one or two bright remnants, hence the very non-uniform
distribution of intensity vs $b$ seen in Figure~\ref{modsvsb}.  This simple
model is ignores correlation of supernova events, which is likely to create
fewer large superbubbles filled with hot gas.

\begin{figure}[tbp]
\includegraphics[width=6in]{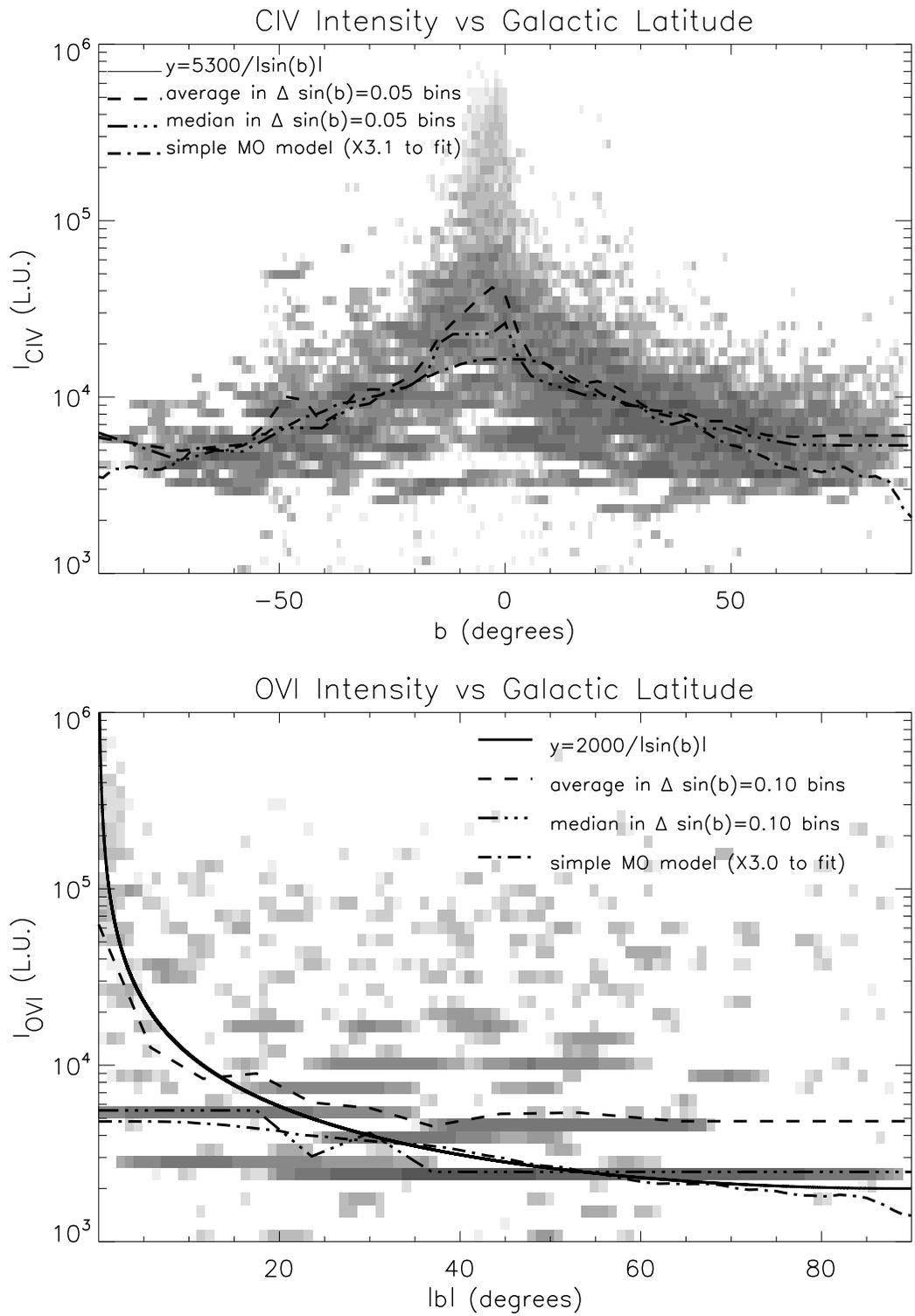}
\caption{Distribution of \CIV\ (top) and \OVI\ (bottom) emission vs galactic
latitude. \label{obsvsb}}
\end{figure}

\section{Structure of Observed Emission}

We show the latitude variation of the emission in Figure~\ref{obsvsb}.  The grey
scale density is proportional to the log of the number of pixels at that $b$
with the observed intensity. The dashed lines are the average emission at that
$b$, and the double-dot-dash line is the median.  The observed emission at 
\CIV\ shows a distinctly bright Galactic plane and faint poles.  The dot-dashed
line is the expected profile of the MO model discussed above.  Because the
sensitivity in the \OVI\ band was lower than that for \CIV, it was necessary to
average larger areas of sky, hence the large horizontal extent of regions in the
\OVI\ plot.  

Overall the median distributions are very consistent with MO model, however the
observed intensities are a factor of 3 higher than for the MO model.  Increasing
the model pressure to 4300 \pressure resolves this, but pushes the neutral gas
filling factor to low levels.  One possible resolution is a MO type
model without pressure equilibrium, with neutral gas at pressures near or below
2500 \pressure, transition temperature gas near 4300 \pressure, and hot gas at
the observed pressures near 10$^{4}$ \pressure.

Many of the high intensity regions near the galactic plane are related to known
supernova remnants.  The Vela SNR and the Cygnus loop SNR are especially
prominent in the FUV (~500k LU). These isolated SNR are more concentrated to
the galactic plane than our simple model would suggest and are visible at
further distances than our uniform dust distribution would indicate.  This can
be anticipated for a clumpy dust distribution.

Some bright region at high-$b$ are related to known superbubbles, especially
Loop 1 and the Orion Eridanus superbubble. 
Many of the small scale intense regions at high-$b$ have not yet been associated
with known structures.  Some of these may be regions where halo gas is cooling.
We note that at high latitudes the average intensity lies up to 3000 LU
(\CIV) or 5000 LU (\OVI) above the MO model expectations, which could indicate
small scale intense regions where \OVI\ outshines \CIV\ as would be expected in
cooling galactic fountain gas.  This could indicate a hot galactic fountain flow of up to 3.5 \solarmassesperyr.

\section{Conclusions}

As expected, each model appears to have some elements of truth. This simple
analysis only includes the two brightest emission lines arising in transition
temperature gas.  More work is being done to incorporate other emission lines
of high stage ions in
the \SPEAR\ bands (\CIII\ \singlet 977, \OIIIb\ \doublet 1661,1667, \SiIV\ 
\doublet 1394,1403, \OIV\ \doublet  1401,1407, \HeII\ \singlet 1640)
into this analysis.  It is our hope that this can confirm the identification of
the source mechanisms in emitting regions.

\begin{theacknowledgments}
This work has been funded by NSF Grant AST-0709347, NASA Grant NAG 5-5355 and by the Korean Ministry of Science and Technology.
\end{theacknowledgments}

\bibliographystyle{aipproc}

\begin{thebibliography}{99}
\bibitem{edelstein06a}Edelstein, J., et al.\ 2006a ApJ, 644, L153
\bibitem{edelstein06b}Edelstein, J., et al.\ 2006b ApJ, 644, L159
\bibitem{korpela06}Korpela, E.~J., et al.\ 2006 ApJ, 644, L163
\bibitem{korpela97}Korpela, E.~J. 1997 Ph.D. Thesis, University of California
\bibitem{mckost77}McKee, C. \& Ostriker, J. 1977 ApJ, 218, 148
\bibitem{slavin93}Slavin, J. \& Cox, D. 1993 ApJ, 417, 187
\bibitem{spitzer90}Spitzer, L.~J. 1990 ARA\&A, 28, 71
\bibitem{chianti}Young, P.~R., Del Zanna, G., Landi, E., Dere, K.~P., Mason, H.~E., \& Landini, M.\ 2003, ApJS, 144, 135
\end{thebibliography}

\end{document}